\documentclass[aps,floats,twocolumn,psfig,epsf,prb,showpacs]{revtex4}
\usepackage{graphicx}

\begin{document}

\let\n=\nu
\let\o=\omega
\let\s=\sigma
\def\np{\n'}
\def\sp{\s'}
\def\EL{E_{L}}
\def\EN{E_N}
\def\ES{E_S}
\def\EM{E_M}
\def\ExN{\mbox{e}^{-\beta \EN}}
\def\ExM{\mbox{e}^{-\beta \EM}}
\def\ExL{\mbox{e}^{-\beta \EL}}
\def\ExS{\mbox{e}^{-\beta \ES}}
\def\c{(c_{\s})}
\def\cd{(c_{\s}^{\dagger})}
\def\cp{(c_{\sp})}
\def\cpd{(c_{\sp}^{\dagger})}

\title{Dynamical vertex approximation --- a step beyond dynamical mean field
theory}
\author{A. Toschi$^a$, A. A. Katanin$^{a,b},$ and K. Held$^a$}

\affiliation{$^a$Max-Planck-Institut f\"ur Festk\"orperforschung, 70569
Stuttgart, Germany\\$^b$Institute of Metal Physics, 620219 Ekaterinburg,
Russia}

\date{Version 8, \today }

\begin{abstract}
We develop a diagrammatic approach with local and nonlocal
self-energy diagrams, constructed from the local irreducible vertex. This approach includes the local correlations of
dynamical mean field theory and
long-range correlations beyond. It allows for
example to describe (para-)magnons and weak
localization effects---in
strongly correlated systems.
As a first application, we study the interplay between
nonlocal antiferromagnetic correlations and the strong
local correlations
emerging in the vicinity of a Mott-Hubbard transition.
\end{abstract}


\maketitle

\section{Introduction}
Strongly correlated electron systems represent both an opportunity and
a challenge for modern physics: An opportunity, since fascinating phenomena
occur such as high-temperature superconductivity in cuprates, ``colossal''
magnetoresistance in manganites, and quantum critical behavior in heavy
Fermion compounds. But at the same time they are a challenge, since the very 
same correlations which are responsible for these phenomena make a theoretical
understanding and hence an experimental optimization of these effects
particularly difficult.

One of the key issues which arises due to strong electronic correlations and
which cannot be described by perturbation theory is the Mott-Hubbard
metal-insulator transition.\cite{MH} In this respect,
dynamical mean field theory (DMFT) \cite{DMFT,DMFT2}
was a big step forward to a more thorough understanding of this transition. 
DMFT becomes exact in the limit of high spatial dimensions 
($d\rightarrow\infty$) and already accounts for a
large (local) part of electronic correlations---the part which
provides for the radical changes upon going from a metal to a
Mott-insulator.
Real physical systems are however one-, two-, or
three-dimensional. Hence, nonlocal correlations, which are neglected in
DMFT, may be of importance.  Corrections of order
 $1/d$  have been considered in Ref.\ \onlinecite{Schiller}, resulting in a 
two-impurity problem, and account
for short-range correlations. There has also
been recent progress to
go beyond DMFT through cluster extensions,\cite{clusterDMFT} which
include correlations within the cluster. These correlations are also
necessarily short-range in nature due to numerical limitations of the cluster 
size.\cite{clusterDMFTFlex}

Often, however, long-range correlations are of vital importance. They are
responsible for a rich variety of phenomena, ranging from magnons and 
screening of the Coulomb interaction to quantum criticality. Long-range 
correlations are also
generally pivotal in the vicinity of phase transitions. The
existing
theories describing long-range (e.g., magnetic) correlations such as the
fluctuation exchange approximation \cite{FLEX}, the 
two-particle self-consistent approximation\cite{vilk}, and the functional
renormalization group \cite{fRG} are restricted to the weak-coupling regime.
For strongly correlated systems, e.g., in the vicinity of a Mott-Hubbard
transition, an extension of DMFT by nonlocal (particularly long-range)
correlations is hence needed.

For
\emph{static} mean-field theories, such corrections have been studied since
decades, e.g., for localized \cite{Heis1z} and itinerant magnets; \cite
{Moriya,MFc} for disordered systems such nonlocal effects
have also been considered.\cite{Janis}
 But there have been only a very few attempts so far to include
long-range correlations beyond dynamical mean field theory: The DMFT 
self-energy was supplemented by an ``external'' $\mathbf{k}$-dependent 
self-energy which describes spin fluctuations in the spin-fermion model \cite
{Sadovskii05} or which stems from the self-consistent renormalization theory,
\cite{Saso} and one might also subsume the GW+DMFT approach \cite{GWDMFT}
here. Let us also note the extended DMFT (E-DMFT \cite{EDMFT}), which
considers the effect of \emph{non-local} interactions on the purely
\emph{local} self-energy.

In this paper, we aim at a systematic diagrammatic extension of DMFT by
long-range
correlations and at an investigation of their effect on the \emph{
non-local} self-energy. Diagrammatically, DMFT corresponds to all 
topologically
distinct, but \emph{local} Feynman diagrams for the self-energy.
On the next level we assume the locality of the fully
irreducible two-particle vertex, and consider all (local or nonlocal)
self-energy diagrams which can be constructed from this vertex.
One might generalize this approach,
requiring locality of the fully irreducible $n$-particle vertex. Then, one
has DMFT for $n=1$, the dynamical vertex approximation
(D$\Gamma$A) for $n=2$, and the exact solution for $%
n\rightarrow \infty$. We think however that the one- and two-particle
levels ($n=1,2$) are the most relevant approximations.
If one is interested in a specific physical problem,
a restriction of D$\Gamma$A
to certain ladder diagrams is reasonable.
In the particle-hole channels the ladder diagrams yield
(para-)magnons  \cite{Moriya} and RPA screening;
in the particle-particle channel the cooperon diagrams
are responsible for attractive pairing interactions and
weak localization effects.  D$\Gamma$A  includes
such ladder
diagrams beyond DMFT, but with the local vertex instead of the bare
interaction so that strong correlations are accounted for.

In this paper we introduce D$\Gamma$A  and apply
it for
studying long-range antiferromagnetic fluctuations in the three-dimensional
Hubbard model. The interplay of these nonlocal
spin
fluctuations with the local DMFT fluctuations is surprising: Close to the
metal-insulator transition the  nonlocal
fluctuations strongly suppress the spectral function,
in contrast to the weak-coupling expectation for
three dimensions.\cite{WC}

The plan of the paper is the following: In Sec. II,
 we introduce the   D$\Gamma$A. Specifically, we derive the
full  D$\Gamma$A scheme based on the parquet equations in
Sec. IIA, and a simplified version based on a ladder
subset of diagrams in Sec. IIB.
The latter yields the most important diagrams
  for the specific problem
considered in this paper, i.e., the
paramagnon fluctuations in the proximity of the AF transition.
We compare this approach
with the $1/d$ expansion in Sec. IIC.
The details concerning the calculation of the local
vertex within an
exact diagonalization impurity solver are
reported in Sec. III
(and in the Appendix).
Results for the
local irreducible vertex and the D$\Gamma$A self-energy and spectral
functions are presented
in Secs. IVA and IVB respectively. Finally we give a
conclusion and discuss the potential of our new method in Sec. V.

\section{Dynamical vertex approximation}

Starting point of our
considerations is the Hubbard model  on a cubic lattice
\begin{equation}
H=-t\sum_{\langle ij\rangle \sigma }c_{i\sigma }^{\dagger }c_{j\sigma
}+U\sum_in_{i\uparrow }n_{i\downarrow }  \label{H}
\end{equation}
where $t$ denotes the hopping amplitude between nearest-neighbors,
$U$ the Coulomb interaction, $c_{i\sigma }^{\dagger }$($c_{i\sigma}$) creates (annihilates) an electron with spin $\sigma$ on site $i$, $n_{i\sigma }\!=\!c_{i\sigma }^{\dagger }c_{i\sigma}$. In the following, we restrict ourselves to the paramagnetic phase with $n$ electrons/site and temperature $T$.

Let us suppose we know the two-particle
vertex
$\Gamma _{\mathbf{kk{^{\prime}}q}}^{\nu \nu ^{\prime } \omega \uparrow\downarrow}$. Then, we can calculate the
(nonlocal) self-energy through the exact relation
(following from the equation  of motion;
see Fig.\ 1a and, e.g., Refs.\ \onlinecite{parquet,Janis2})
\begin{eqnarray}
\Sigma _{\mathbf{k},\nu} \!=\!U \frac{n}{2}\! -T^2U\!\sum_{\stackrel{\scriptstyle \nu{^{\prime }}\omega}{\mathbf{k^{\prime}\mathbf{q}}}}\Gamma _{\mathbf{k
k{^{\prime}}q}}^{\nu \nu ^{\prime } \omega\uparrow\downarrow }G_{\mathbf{k}^{\prime }\mathbf{\!+q},\nu
^{\prime }\!+\omega }G_{\mathbf{k}^{\prime }\!,\nu ^{^{\prime }}}G_{\mathbf{k\!+q}%
,\nu +\omega }
\label{Eq:DGA}
\end{eqnarray}
where $G_{\bf{k},\nu}=(i \nu_n-\epsilon_{\bf{k}}+\mu-\Sigma_{\bf{k},\nu})^{-1}$ is the nonlocal Green function,
$\epsilon_{\mathbf{k}}=-2t \sum_{\alpha =x,y,z} \cos{k_\alpha}$  the bare electronic dispersion, and $\mu$ the
electronic chemical potential. Generally, $\Gamma _{\mathbf{kk{^{\prime}}q}}^{\nu \nu ^{\prime }
\omega\uparrow\downarrow }$ can be expressed diagrammatically, e.g., by taking the  fully two-, three- and more
particle irreducible local vertices as building blocks and connecting these blocks by local and nonlocal Green functions.
\begin{figure}[tb]
\includegraphics[width=8.7cm]{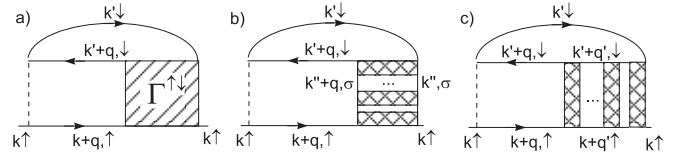}
\caption{a) From the reducible vertex we directly obtain the
self-energy. b) and c) The two particle-hole
channels contributing to the self-energy (longitudinal and transversal)
in the ladder approximation. Instead of the
bare interaction, ladder diagrams are constructed from
the irreducible local vertices (crosshatched) in D$\Gamma$A.}
 \vspace{-3mm}
\label{fig1}
\end{figure}

\subsection{Parquet equations}
\label{Sec:Parquet}
\begin{figure*}[tbh]
\begin{minipage}{12cm}
{\includegraphics[width=12.cm]{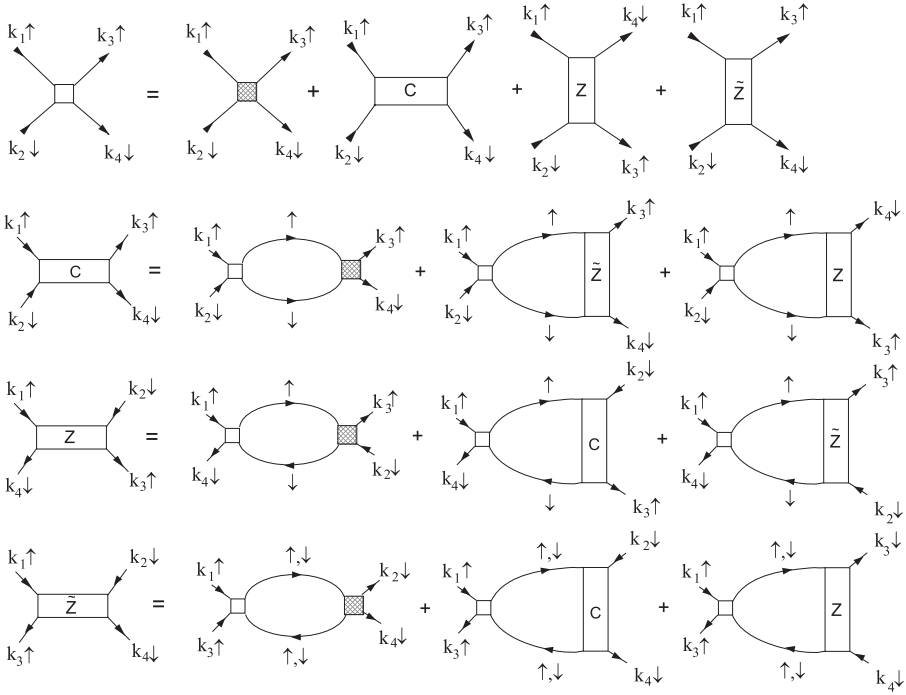}}
\end{minipage}\hfill
\begin{minipage}{4.3cm}
\caption{Graphical notation of the parquet Eqs.\ (\ref{Eq:Parquet}), (\ref{Eq:Parquet2a}), (\ref{Eq:Parquet2b}), and  (\ref{Eq:Parquet2c}).
$C$, $Z$, and $\tilde{Z}$ denote the contributions of the
particle-particle, particle-hole and interaction channels to the
non-local vertex $\Gamma _{\mathbf{k_1,k_4,k_3-k_1}}^{\nu_1, \nu_4,\nu_3-\nu_1,  \uparrow\downarrow}$ (the
momentum and frequency $\mathbf{k}_2,\nu_2$ are determined by conservation laws). \label{Fig:parquet}} \end{minipage}
\end{figure*}

In the D$\Gamma$A, we restrict
ourselves to the local fully-irreducible two-particle vertices
$\Gamma^{\rm loc}_{\rm fir}$. From these building blocks,
the reducible vertices
 $\Gamma _{\mathbf{kk{^{\prime}}q}}^{\nu \nu ^{\prime } \omega \uparrow\downarrow}$
can be obtained through the self-consistent solution of the
parquet equations.\cite{Dzy,parquet} Representing this vertex as a sum of contributions of different channels, one has
\cite{Dzy,parquet,Janis2}
\begin{eqnarray}
\Gamma _{\mathbf{k
k{^{\prime}}q}}^{\nu \nu ^{\prime } \omega \uparrow\downarrow}&=&\Gamma^{\nu\nu^{\prime } \omega, \uparrow\downarrow}_{\text{fir,loc}}
+C _{\mathbf{k k{^{\prime}}q}}^{\nu \nu ^{\prime } \omega}+
Z _{\mathbf{k k{^{\prime}}q}}^{\nu \nu ^{\prime } \omega}+
\tilde{Z} _{\mathbf{k
k{^{\prime}}q}}^{\nu \nu ^{\prime } \omega} \label{Eq:Parquet}\\
\Gamma _{\mathbf{k
k{^{\prime}}q}}^{\nu \nu ^{\prime } \omega \uparrow\uparrow}&=&\Gamma _{\mathbf{k
k{^{\prime}}q}}^{\nu \nu ^{\prime } \omega \uparrow\downarrow}-\bar{\Gamma} _{\mathbf{k
k{^{\prime}}q}}^{\nu \nu ^{\prime } \omega \uparrow\downarrow}.
\end{eqnarray}
Here,
$
\bar{\Gamma} _{\mathbf{k
k{^{\prime}}q}}^{\nu \nu ^{\prime } \omega \uparrow\downarrow}=
\Gamma _{\mathbf{k,
k+q,k-k{^{\prime}}}}^{\nu, \nu+\omega,\nu-\nu^{\prime }  \uparrow\downarrow}$
and the contribution of the three channels can be written in the following form
(see Fig.\ \ref{Fig:parquet})
\begin{eqnarray}
C _{\mathbf{k k{^{\prime}}q}}^{\nu \nu ^{\prime } \omega}&=&
\Gamma^{ \uparrow\downarrow} *G*G*(\Gamma^{ \uparrow\downarrow}_{\text{fir,loc}}+Z+\tilde{Z})\label{Eq:Parquet2a}\\
Z _{\mathbf{k k{^{\prime}}q}}^{\nu \nu ^{\prime } \omega}&=&
\Gamma^{ \uparrow\downarrow} *G*G*(\Gamma^{ \uparrow\downarrow}_{\text{fir,loc}}+C+\tilde{Z})\label{Eq:Parquet2b}\\
\tilde{Z} _{\mathbf{k k{^{\prime}}q}}^{\nu \nu ^{\prime } \omega}&=&
(\Gamma^{ \uparrow\uparrow}+\Gamma^{ \uparrow\downarrow}) *G * G*(\Gamma^{
\uparrow\downarrow}_{\text{fir,loc}}+C+Z)\nonumber\\
&-&
\Gamma^{ \uparrow\downarrow} *G * G*(\bar{\Gamma}^{
\uparrow\downarrow}_{\text{fir,loc}}+\bar{C}+\bar{Z})
\label{Eq:Parquet2c}
\end{eqnarray}
where $*$ stands for multiplication and summation over
the momenta and frequencies
 given in Fig.\ \ref{Fig:parquet}.

For determining the fully irreducible local 
vertex $\Gamma^{\rm loc}_{\rm fir}$, which is constructed from purely local 
Feynman diagrams only, we resort to the Anderson impurity model 
(AIM)\cite{AIM}. In fact, the AIM has only one interacting site, so 
it yields the same local diagrams -and hence the same irreducible vertex- 
provided that the local Green function is identical.
Hence a practical way to obtain  $\Gamma^{\rm loc}_{\rm fir}$ is through the 
(e.g., numerical) solution of the AIM. 
Starting point can be the (local) spin and charge susceptibility
 of the AIM \begin{equation}
\chi_{\text{s(c),loc}}^{\nu \nu' \omega}=\chi_{\text{loc}}^{\nu \nu' \omega,\uparrow\uparrow} \hspace{-2mm} {\scriptsize {\tiny  \begin{array}{cc} +\\(-) \end{array}} }
\chi_{\text{loc}}^{\nu \nu' \omega,\uparrow\downarrow} \label{hisc}
\end{equation}
from which
we can obtain the full (reducible) local vertex
$\Gamma_{s(c), loc}$ and the irreducible local
vertices in the  spin-, charge- and particle-particle channel ($\Gamma_{\text{s,ir}}$, $\Gamma_{\text{c,ir}}$
and $\Gamma_{\text{pp,ir}}$, respectively) via the standard relations:
\begin{eqnarray}
\chi_{\text{s(c),loc}}^{\nu \nu' \omega}&=&\chi_{0\omega, \text{loc}}^{\nu}\delta_{\nu \nu'}+\chi_{0\omega, \text{loc}}^{\nu}
\Gamma_{\text{s(c),loc}}^{\nu \nu' \omega}\chi_{0\omega, \text{loc}}^{\nu'}\\
\Gamma _{{s}{(c)},\text{loc}}^{\nu\nu ^{\prime } \omega }&=&[(\Gamma _{s (c),\text{ir}}^{\nu
\nu^{\prime } \omega })^{-1}-\chi _{0 \omega, \rm{loc}}^{\nu ^{\prime }}\delta _{\nu \nu ^{^{\prime }}}]^{-1},\\
\Gamma_{\text{pp,loc}}^{\nu\nu^{\prime } \omega,\uparrow\downarrow }&=&[(\Gamma _{\text{pp,ir}}^{\nu, \nu ^{\prime },
\tilde{\nu}-\nu })_{\nu,\tilde{\nu}}^{-1}-\Pi _{0 \tilde{\nu}+\nu^{\prime}, \rm{loc}}^{\tilde{\nu} }\delta _{ \nu
\tilde{\nu} }]_{\tilde{\nu}=\nu+\omega}^{-1},
\end{eqnarray}
where
\begin{eqnarray} \chi _{0\omega,\text{loc}}^{\nu
^{\prime }}=-T G_{\text{loc}}(\nu ^{\prime })G_{\text{loc}}(\omega+\nu ^{\prime }), \nonumber \\ \Pi
_{0\omega,\text{loc}}^{\nu ^{\prime }}=T G_{\text{loc}}(\nu ^{\prime })G_{\text{loc}}(\omega-\nu ^{\prime }).
\label{hiPi}
\end{eqnarray}
From these vertices we can in turn calculate the
 fully irreducible vertex as
\begin{eqnarray}
\Gamma^{\nu\nu^{\prime } \omega, \uparrow\downarrow}_{\text{fir,loc}}=\frac{1}{2}(\Gamma
_{s,\text{ir}}^{\nu\nu^{\prime } \omega }-\Gamma _{c,\text{ir}}^{\nu\nu^{\prime } \omega })+\Gamma
_{s,\text{ir}}^{\nu,\nu+\omega,\nu^{\prime}-\nu }\nonumber\\+ \Gamma _{\text{pp,ir}}^{\nu\nu^{\prime } \omega
}-2\Gamma _{\text{loc}}^{\nu\nu^{\prime } \omega \uparrow\downarrow}.
\label{Eq:fir}
\end{eqnarray}

\subsection{Ladder approximation}

In this paper, we are particularly interested in paramagnon contributions
affecting the self energy in the vicinity of the antiferromagnetic phase. Hence, as discussed in the
introduction and with some justification in $1/d$ (see Sec. II C), we
restrict ourselves to the ladder subset of the
parquet diagrams in the two particle-hole channels shown in Fig.\ 1b and 1c.
These diagrams can be derived from the general parquet set of diagrams of 
Fig.\ 2, supposing the locality of $C,Z, \tilde{Z}$ in the r.h.s. of Eqs. (\ref{Eq:Parquet2a})-(\ref{Eq:Parquet2c}). 
Expressing the ladder diagrams
through the vertices in the  spin (s) and charge (c) channels $\Gamma _{s(c),\mathbf{q}}^{\nu\nu^{\prime } \omega }$,
which depend on the momentum transferred $\bf{q}$ only,
the  sum of the vertices of Fig.\ 1b and 1c is obtained as
\begin{eqnarray}
\Gamma _{\mathbf{k
k{^{\prime}}q}}^{\nu \nu ^{\prime } \omega,\uparrow \downarrow} &=&\frac 12(\Gamma
_{s,\mathbf{q}}^{\nu \nu ^{\prime } \omega }-\Gamma _{c,\mathbf{q}%
}^{\nu \nu ^{\prime } \omega })+\Gamma _{s,\mathbf{k^{\prime }-k}}%
^{\nu ,\nu + \omega ,\nu ^{\prime }-\nu  } \nonumber \\
&&-\frac 12
(\Gamma _{s,\text{loc}}^{\nu\nu  ^{\prime }\omega }
-\Gamma _{c,%
\text{loc}}^{\nu \nu ^{\prime } \omega }).
\label{Eq:SimpDGA}
\end{eqnarray}
Here, the first two terms of Eq.\ (\ref{Eq:SimpDGA})
describe the
longitudinal and transverse paramagnons in Fig.\ 1b and 1c, respectively, and
the last term
subtracts the double-counted local contribution.
Note that the nonlocal contribution of the particle-particle
channel to the self-energy, which is not relevant
near magnetic instabilities, has been neglected here.

 The quantities on the right
hand side of Eq. (\ref{Eq:SimpDGA}) are  calculated  from the
local vertex $\Gamma _{s(c),\text{ir}}^{\nu\nu^{\prime }\omega }$,
irreducible in the spin (charge) channel, via
\begin{eqnarray}
\Gamma _{s(c),\mathbf{q}}^{\nu\nu^{\prime } \omega }=[(\Gamma _{s (c),\text{ir}}^{\nu
\nu^{\prime } \omega })^{-1}-\chi _{0\mathbf{q} \omega }^{\nu ^{\prime }}\delta _{\nu \nu
^{^{\prime }}}]^{-1},
\label{Eq:Gamma}
\end{eqnarray}
where
$\chi _{0\mathbf{q}\omega}^{\nu ^{\prime }}=-T\sum_{\mathbf{k}}G_{\mathbf{k},\nu ^{\prime } }G_{\mathbf{k}+%
\mathbf{q},\nu ^{\prime } +\omega }$  with
$G_{\mathbf{k},\nu}=[i\nu-\epsilon_{\mathbf{k}}+\mu - \Sigma_{\text{loc}}(\nu)]^{-1}$, $\Sigma_{\text{loc}} $ being the local (DMFT) self-energy. 
Note that contrary to the full parquet solution in Sec. \ref{Sec:Parquet} the
self-energy of the internal Green functions is considered purely local in accordance with the assumption of the
locality of the vertex $\Gamma _{s (c),\text{ir}}^{\nu \nu^{\prime } \omega }$. The results of this non-self-consistent
approach are expected to be close to those of the self-consistent one, due to the cancellations between
(self-consistent) non-local self-energy and corresponding corrections to the vertex $\Gamma _{s (c),\text{ir}}$, cf. Ref. \onlinecite{DiCastro}.

Substituting Eq.\ (\ref{Eq:SimpDGA}) into Eq.\ (\ref{Eq:DGA}), we obtain after a shift of the momenta and frequencies
\begin{eqnarray}
\Sigma _{\mathbf{k},\nu } &=&
\frac{1}{2}{Un}+\frac{1}{2}TU\sum\limits_{\nu ^{\prime }\omega ,\mathbf{q}%
}\chi _{0\mathbf{q}\omega }^{\nu ^{\prime }}\left( 3\Gamma _{s,%
\mathbf{q}}^{\nu\nu ^{\prime } \omega }-\Gamma _{c,\mathbf{q}}^{\nu
\nu ^{\prime } \omega }\right.   \nonumber \\
&&\left. +\Gamma _{c,\text{loc}}^{\nu\nu ^{\prime } \omega }-
\Gamma _{s,\text{loc}}^{\nu\nu ^{\prime } \omega }\right) G_{\mathbf{k+q}%
,\nu +\omega }.
\label{Eq:final}
\end{eqnarray}
Eq.\ (\ref{Eq:final}) reduces to the DMFT self-energy if
the nonlocal quantities are replaced by local ones.
But beyond that, it describes the nonlocal ladder diagrams of Fig.\ 1b and 1c.


\subsection{Comparison to the $1/d$ expansion}

\begin{figure}[t!]
\includegraphics[width=8.5cm]{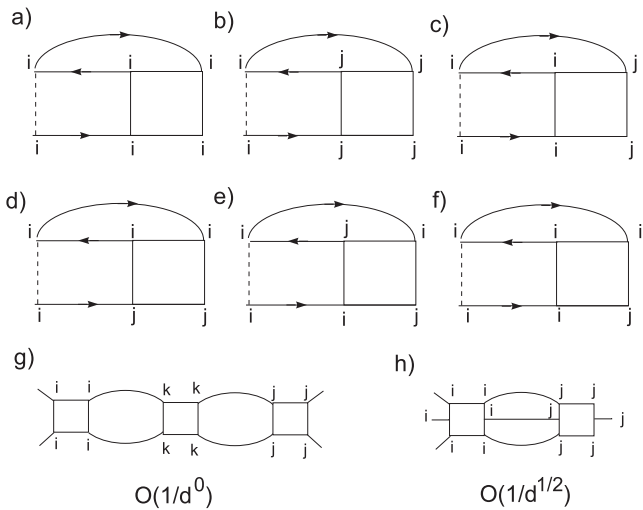}
\caption{Diagrams for the self-energy in terms of the vertex to order $1/d^0$,
[diagram a), i.e., DMFT] and $1/d^{1/2}$ [diagrams b)-f)]. The ladder
diagram g) show how diagrams b)-e) can be constructed from a fully irreducible local two-particle vertex. In contrast, for diagram f) a local three-particle vertex is needed, see h). Hence the contribution  f) is the only one not included in the D$\Gamma$A.}
\label{Fig:1overd}
\end{figure}

Let us compare the result Eq.\ (\ref{Eq:final}) to that of the $1/d$ expansion.
While the DMFT self-energy contains the local Green functions and vertex only (Fig. 3a), the
leading nonlocal corrections to the self-energy are proportional to $O(1/d^{1/2})$ and the consideration of the
diagrams containing two different sites is sufficient at this order.\cite{Schiller} The possible types of these
diagrams for the self-energy are shown in Fig. \ref{Fig:1overd}b-\ref{Fig:1overd}f. The first type of corrections
(Fig. \ref{Fig:1overd}b) involve the non-local Green functions only. The second type of corrections (Figs.
\ref{Fig:1overd}c-\ref{Fig:1overd}e) contains the non-local vertices with two legs at $i$ and two at $j$ sites. To
leading (zeroth) order in $1/d$ these vertices can be expressed as a ladder of local vertices connected by
the non-local Green functions, as in Fig.  \ref{Fig:1overd}g. The contributions of Figs. \ref{Fig:1overd}b-\ref{Fig:1overd}e
are hence included in the D$\Gamma$A with the ladder approximation, i.e., in Eq. (\ref{Eq:final}).
The last type of $1/d^{1/2}$ corrections to the self-energy (Fig. \ref{Fig:1overd}f) involve the three-particle local
vertex (Fig. \ref{Fig:1overd}h) (and are of order $(U/t)^5$). According to the classification of the introduction,
these corrections should be taken into account on
 the next level of approximation
beyond  D$\Gamma$A   and  are the only $1/d^{1/2}$ corrections neglected in  D$\Gamma$A.
Therefore, the D$\Gamma$A reproduces
correctly
the leading $1/d$ correction to the self-energy with the three-particle local
 vertex neglected.

\section{Calculation of the local four-point vertex}

The calculation of the self-energy (\ref{Eq:final})
requires the knowledge of
the local vertex,  either fully irreducible---for
the general scheme---, or
irreducible in the  spin (charge) channel---for
the ladder diagrams of Fig.\ 1b and 1c
[Eq.\ (\ref{Eq:final})].
As already noted,
this  local vertex can be obtained numerically from
the Anderson impurity
model. For obtaining the four-point vertex
$\Gamma _{s(c),\text{loc}}^{\nu ^{\prime }\nu \omega }$,
we need to
calculate the AIM susceptibility for three
Matsubara frequencies\cite{note_RMP}
\begin{eqnarray}
\label{Eq:chiloc}
\chi_{loc\,}^{\nu\nu'\omega {\sigma}{\sigma}^{'}}  &  = & T^2
\int_0^{1/T} d\tau_1\, d\tau_2 \, d\tau_3 \;
\mbox{e}^{-i\tau_1\nu}\, \mbox{e}^{i\tau_2(\nu+\omega)} \,\mbox{e}^{-i\tau_3(
\nu'+\omega)} \nonumber  \\ 
& \times & \left[ \langle T_{\tau} \; c_{i\sigma}^{\dagger}(\tau_1)c_{i\sigma}(\tau_2)c_{i\sigma'}^{\dagger}(\tau_3)c_{i\sigma'}(0) \rangle \right. \nonumber  \\ 
& -  & \left.  \langle T_{\tau} \; c_{i\sigma}^{\dagger}(\tau_1) c_{i\sigma}(\tau_2) \rangle \langle T_{\tau}  c_{i\sigma'}^{\dagger}(\tau_3)c_{i\sigma'}(0) \rangle  \right] .
\end{eqnarray}
where $\nu,\nu'$  and $\omega$ are  the two fermionic and the
bosonic  (transferred) Matsubara frequency, respectively;
$\langle T_{\tau} \cdots \rangle$ indicates the thermal
expectation
value  of the time-ordered operators and the last term represents the 
non-connected contributions. With a spin (anti-) symmetrization (\ref{hisc})
we obtain the corresponding  charge and spin
susceptibilities $\chi_{s(c), \text{loc}}^{\nu\nu'\omega}$.
From these, we
can either determine the fully irreducible local vertex and
through the parquet equations the reducible vertex and
the self energy along the lines of Eqs.\ (\ref{Eq:fir}),  (\ref{Eq:Parquet}), and  (\ref{Eq:DGA});
or we can directly calculate the particle-hole
ladders along the lines of Eqs.\ (\ref{Eq:SimpDGA}) and (\ref{Eq:final}).
We implement the latter by (i) solving the DMFT equations using
  exact diagonalization (ED), (ii) calculating
via Eq.\ (\ref{Eq:chiloc}) the local vertices, and
(iii) constructing from these through Eq.\ (\ref{Eq:final})
the D$\Gamma$A self-energy.
In principle, this $k-$dependent self-energy yields new Green functions and 
a new vertex. However, because of the ladder approximation, we do not perform 
such a self-consistent calculation here.

Within ED, the calculation of
 $\chi_{\text{loc}}$ [Eq.\ (\ref{Eq:chiloc})] is
straightforwardly (albeit lengthy) performed
 by resorting to its Lehmann
representation, whose explicit expression is reported in
the Appendix.  The Lehmann representation
of $\chi_{\text{loc}}$ requires  four summations over
all the Hilbert states of the discretized AIM
 (compared with only two summations for evaluating the
local Green function). This is the higher
computational cost of D$\Gamma$A compared with DMFT.
By performing a parallel computation of $\chi_{loc}$,
we were able to calculate AIMs with $N_s\!=\!5$
sites and evaluate  $\chi_{loc}$ for the lowest
$N_{max} \! = \! 20$ (or, in some cases, $25$) Matsubara frequencies.
This turned out to be sufficient for getting a stable
analytic continuation of Eq.\ (\ref{Eq:final}), using the Pad\'e algorithm.
For the momentum summation of Eq.\ (\ref{Eq:final})
we have used $N_k=96$ points for each directions.

\section{Results}

\subsection{Local vertex close to a Mott-Hubbard transition}

\begin{figure}[t]
\includegraphics[width=8.8cm]{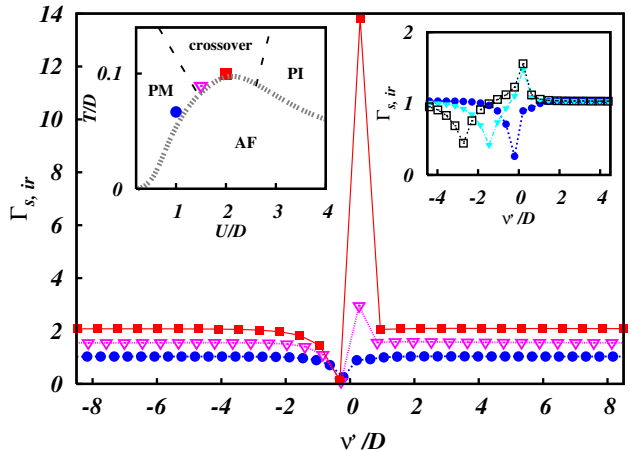}
\caption{(Color online) Dependence of the local vertex
${\Gamma_{s,\rm{ir}}^{\nu \! =\! \pi \, T,\nu^{\prime}\!,\omega=0 }}$
on the incoming Fermionic frequency
${\nu^{\prime }}$,
for the three different values of  $U$ and $T$ indicated
as symbols in the left inset, which shows
the DMFT phase diagram with paramagnetic metallic (PM),
insulating (PI), and antiferromagnetic (AF) phase.
Right inset: Same as main panel
but at fixed $U=1D$ and  at different $\omega$'s  (circles: $\omega\!=\!0$;
 triangles:  $\omega\!=\!6\pi T$; squares: $\omega\!=\!12\pi T$).}
\label{Fig:Gamma}
\end{figure}

\begin{figure*}[thb]
\hspace{-5mm}
\begin{minipage}{12.5cm}
{\includegraphics[width=12.5cm]{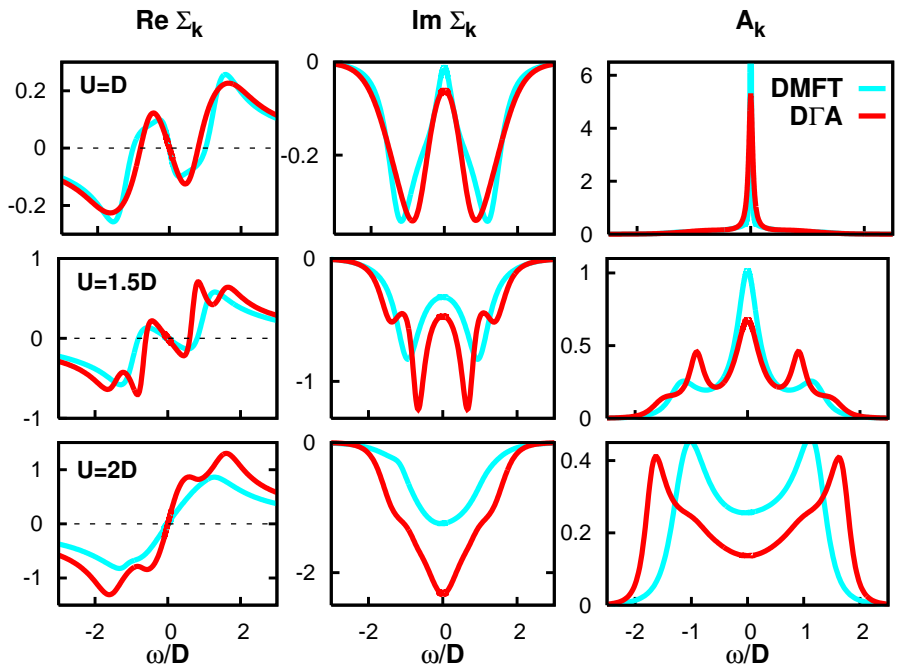}}
\end{minipage}\hfill
\begin{minipage}{4.3cm}
\caption{(Color online) Self energy (left: real, middle: imaginary part) and spectral function  (right)  at ${\bf k}\!=\!(\pi/2,\pi/2,\pi/2)$,
$U\!=\!1D$ (top),  $1.5D$ (central), and $2D$ (bottom) and the same $T$'s of Fig.\ \ref{Fig:Gamma}.
Compared to DMFT (light blue line)
quasiparticles are damped through scattering
at nonlocal spin fluctuations in D$\Gamma$A (dark red line),
and the system is more insulating.
 At $U\!=\!1.5D$ and $2D$,
these effects are drastically enhanced because
of strong local correlations reflected in
$\Gamma_{\rm{s,ir}}$
\label{Fig:Sk}}
\end{minipage}
\end{figure*}

First, we discuss our results for the local irreducible vertex.
Since we are mainly interested in
 long-range antiferromagnetic fluctuations, we consider
 temperatures  slightly above the DMFT N\'eel temperature
(see the inset of Fig.\ \ref{Fig:Gamma}).
In this case the largest contribution  to the nonlocal
part of the D$\Gamma$A
self-energy  stems from the terms
 of Eq.\ (\ref{Eq:final})  proportional to the
local vertex irreducible in the spin channel, particularly with zero bosonic
frequency.

In Fig. \ref{Fig:Gamma},  we show the ${\Gamma_{s,\text{%
ir}}^{\nu \nu ^{\prime }, \omega=0 }}$ as a function of $\nu'$ (with
$\nu$ fixed to its lowest value $\pi T$)  at half-filling for three
different values of the Hubbard interaction
($U=1D$, $U=1.5 D$ and $U=2D$,where $D=2\sqrt{6}\, t$ is twice the variance of
the non-interacting density of states) 
and temperature ($T=0.067 D$, $0.089D$ and $0.1D$).
This choice allows us to highlight the
remarkable differences occurring when moving from the metallic regime to the
crossover region of the DMFT phase-diagram of the Hubbard
model.
In particular, we note that
$\Gamma_{s,ir}$ correctly
approaches the corresponding value of the bare interaction $U$ for large
$\nu'$. On the
other hand, at small $\nu'$ we observe a radically different behavior of the
local vertex depending on the $U$ value: At $U=1D$,
$\Gamma_{s,ir}$ displays
a smooth minimum in the region of small $\nu'$,
while at $U=1.5D$  (and even more at $U=2D$) a very pronounced  maximum
of $\Gamma_{s,ir}$ appears at $\nu'=\nu$.  While the behavior of
$\Gamma_{s,ir}$ at small $U$ can be
easily interpreted as the screening of the bare interaction, typical of the
metallic phase, the huge maximum of
$\Gamma_{s,ir}$ at larger $U$ stems from particle-hole fluctuations in
the vicinity of the metal-insulator transition, as discussed 
in  Ref. \onlinecite{Janis2}.

\subsection{D$\Gamma$A self-energy and spectral function}

The  striking behavior of  $\Gamma_{s,ir}$
has consequences for the  D$\Gamma$A self-energy on the real axis
and spectral function, presented in Fig. \ref{Fig:Sk}
 for the same three
different $U$'s and $T$'s of
 Fig.\ \ref{Fig:Gamma} (left inset).

At $U=1D$, i.e.,
in the metallic regime of the phase diagram,
DMFT shows a quasiparticle peak which is only weakly damped
for ${\bf k}$-vectors  on the Fermi surface.
In D$\Gamma$A, the quasiparticle
scatters at nonlocal antiferromagnetic
fluctuations, resulting in a
broadening  of the
quasiparticle with a now significant damping
given by $\text{Im} \Sigma_{\bf k}(0)$.
For this $U$ value, the distinct features of
 $\Gamma_{s,ir}$  are not yet particularly
pronounced in Fig.\ \ref{Fig:Gamma}.
It is the strongly enhanced antiferromagnetic
susceptibility close to the N\'eel temperature
which leads to this damping in Eq.\ (\ref{Eq:final}).
Stronger damping effects can be observed, studying the Hubbard model 
in $d=2$, e.g.  by means of the cluster extensions of DMFT\cite{clusterDMFT}, 
which predict a pseudogap opening at low temperatures. Actually, a stronger 
damping in $d=2$ than in the three dimensional case considered here 
is expected from weak coupling perturbation theory. Unfortunately, to our
 knowledge, no cluster-DMFT calculation has been performed for the case of 
$d=3$, since it poses severe constraints on the cluster size.

At $U=2D$, the stronger electronic correlation reflects in more pronounced 
 changes of the spectral function. Now, it is the huge $\Gamma_{s,ir}$ of 
Fig.\ \ref{Fig:Gamma}
which strongly suppresses  the spectral weight at the Fermi level.
 This weight is transferred to the Hubbard  subbands,
which get  some additional structure as an effect of magnetic fluctuations.
The nonlocal fluctuations result
in a much more insulating solution, albeit the Green function is
still non-local\cite{note_proof}.

We emphasize that the mechanism
of spectral weight suppression
at $U\!=\!2D$
is  very different from
that in the weak-coupling regime (Ref.\  \onlinecite{WC} and
our $U\!=\!1D$ results),
 where
long-range magnetic fluctuations in the immediate
vicinity of the magnetic phase transition
play the key role.
In contrast at $U\!=\!2D$,
already relatively short-range
spin fluctuations are important
because of the
strong correlations reflected in the enhanced
$\Gamma_{s,ir}$. The spectral weight suppression
is therefore also quite robust upon increasing $T$, i.e.,
upon going further away from the antiferromagnetic transition
(not shown).

At $U\!=\!1.5D$, we have something in between the two cases discussed 
above: The vertex is already enhanced, but
long-range antiferromagnetic fluctuations
are still essential.
This leads to a suppression of the quasiparticle weight
 and structured Hubbard bands.
Altogether this shows that scattering at nonlocal fluctuations
close to a strongly correlated antiferromagnet
is very different from that in the vicinity
of a weakly correlated Slater antiferromagnet.


\section{Conclusions}
We developed the dynamical vertex approximation
(D$\Gamma$A)
based on the assumption of the locality of the
irreducible vertex. For the half-filled
three-dimensional Hubbard model, we found that
the   local vertex (irreducible in the
particle-hole spin channel) strongly depends
on all three frequencies;
it is hugely enhanced at some particular
frequencies for large $U$.
These  strong local correlations
entail similarly  strong
nonlocal  fluctuations.
The scattering of the quasiparticles
 at these nonlocal spin fluctuations, in turn,
drastically reduces their life times;
spectral weight is transferred to the
Hubbard bands which develop some additional structure.
These nonlocal effects of strong electronic correlations
are very different from those
at weak coupling.
It is a strength of D$\Gamma$A
to reveal them.

Including long-range correlations,
 D$\Gamma$A opens the door to study a wide variety
of physical phenomena, previously described only
for weakly correlated systems, such as
magnons in strongly correlated (anti-)ferromagnets\cite{note_af},
the interplay of weak localization effects and
strong electron interactions, and vertex corrections
to the RPA screening.
A self-consistent realization of the approach
might also allow us to study how nonlocal fluctuations
suppress magnetic long-range ordering,
whether antiferromagnetic
fluctuations  in the vicinity of the metal-insulator transition
result in unconventional superconductivity,
and how physical quantities change in the vicinity
of ferro- and antiferromagnetic quantum critical points.

\vspace{3mm}

{\it{Acknowledgments.}}
We thank  W.\ Metzner, M.\ Capone, C.\ Castellani,
G.\ Sangiovanni, and  R.\ Arita for stimulating discussions;
we are indebted to M. Capone also for providing
the DMFT(ED) code which has served as a starting point.
This work was supported by
the  Deutsche Forschungsgemeinschaft
 through the Emmy-Noether program (AT,KH) and 436 RUS 113/723/0-2 (AK)
and by the Russian Basic Research Foundation through Grant No.~747.2003.2 (AK).

\vspace{2mm}

{\it{Note added.}} During the completion of our paper, we  learned about
a related study, Ref.\ \onlinecite{Kusunose}. Another related paper, Ref.\
\onlinecite{Jarrnew}, has appeared just after our pre-print.

\appendix*

\section{ \hspace{4mm} Lehmann representation \hspace{8mm}
of the local susceptibility}

In this Appendix we report the explicit expression of the Lehmann
representation for the local (and spin-dependent) susceptibility, which
is necessary for performing the calculation of the
basic ``brick'' of the D$\Gamma$A  (i.e., the local four point vertex)
within the ED algorithm (see Sec. III).

We start with the evaluation of the T-ordered product appearing in the
definition of the (spin-dependent)  local
susceptibility in Eq. (\ref{Eq:chiloc}):

\begin{eqnarray}
\label{Eq:chiTtau}
\tilde{\chi}_{\text{loc}\,}^{\nu\nu'\omega {\sigma}{\sigma}^{'}} = \, T^2
\int_0^{\beta} d\tau_1\, d\tau_2 \, d\tau_3 \;
\mbox{e}^{-i\tau_1\nu}\, \mbox{e}^{+i\tau_2(\nu+\omega)} \,\mbox{e}^{-i\tau_3(
\nu'+\omega)} \nonumber \\
\times \; \, \langle T_{\tau} \; c_{i\sigma}^{\dagger}(\tau_1)c_{i\sigma}(\tau_2)c_{i\sigma'}^{\dagger}(\tau_3)c_{i\sigma'}(0) \rangle \nonumber \\ \vspace{2mm} \nonumber \\
 = \, T^2 \int_0^{\beta}d\tau_1 \left[ \int_0^{\tau_1}  d\tau_2 \left( \int_0^{\tau_2} d\tau_3 -  \int_{\tau_2}^{\tau_1} d\tau_3 +\int_{\tau_1}^{\beta}d\tau_3\right)\right. \nonumber \\
-  \left. \int_{\tau_1}^{\beta}  d\tau_2 \left( \int_{0}^{\tau_1}
d\tau_3-\int_{\tau_1}^{\tau_2} d\tau_3 +\int_{\tau_2}^{\beta}d\tau_3\right)\right] \hspace{8mm}  \nonumber \vspace{10mm} \\ \vspace{10mm}
\times \; \, \mbox{e}^{-i\nu(\tau_1-\tau_2)}\, \mbox{e}^{i\omega (\tau_2-\tau_3)} \,\mbox{e}^{-i \nu'\tau_3}\; \; \,  \langle \; \,  \cdots \; \, \rangle
  \hspace{14mm}  \nonumber \vspace{15mm}\\ \vspace{2mm} \nonumber \\ \hspace{-20mm} = \frac{T^2}{Z}\; (\chi_{\text{loc}}^{123} + \chi_{\text{loc}}^{132} + \chi_{\text{loc}}^{312}+\chi_{\text{loc}}^{213}+\chi_{\text{loc}}^{231}+\chi_{\text{loc}}^{321}) \hspace{10mm}
\end{eqnarray}
where $\beta=1/T$, $Z$ is the partition function and  with 
$\langle \; \, \cdots \; \, \rangle$ we
indicate the thermal and quantum average of the four fermionic operators
in the r.h.s. of the first line of the equation, which  are already ordered
 in terms of decreasing times (with no further sign change). 
The six different contributions $\chi_{\text{loc}}^{123}, \;
\chi_{\text{loc}}^{132}, \, \cdots$ appearing in the last line of
Eq. (\ref{Eq:chiTtau}) reflect the six different ways of
arranging the order of the three Matsubara times $\tau_1$, $\tau_2$ and
$\tau_3$ in the time integral of Eq. (\ref{Eq:chiloc}).
$\chi_{\text{loc}}^{123}, \; \chi_{\text{loc}}^{132}, \, \cdots$  can be explicitly
expressed in a very convenient way for ED scheme, i.e.,
in terms of the eigenenergies $E_N$ and the matrix elements
$\langle N|c_{\sigma}^{(\dagger)} |M \rangle = (c_{\sigma}^{(\dagger)})_{NM}$,
 of the associated  AIM, through the standard Lehmann representation.
The evaluation of the six time integrals in Eq. (\ref{Eq:chiTtau})
is straightforward- albeit lengthy, and yields the following results:

\begin{widetext}
\begin{eqnarray}
\label{Eq:chi123}
\chi_{\text{loc}}^{123}&  = & \sum_{N,M,L,S}\frac{-1}{i(\np+\o)-\EL+\ES}\left[
\frac{1}{i(\n-\np)+\EM-\ES}
 \left(\frac{\ExN+\ExS}{i\np-\EN+\ES}-\frac{\ExM+\ExN}{i\n-\EN+\EM}\right) \nonumber \right. \\ & - & \left. \frac{1}{i(\n+\o)+\EM-\EL}
\left(\frac{\ExL-\ExN}{i\o+\EN-\EL}-\frac{\ExM+\ExN}{i\nu-\EN+\EM}\right)\right]\times\cd_{NM}\c_{ML}\cpd_{LS}\cp_{SN}
\end{eqnarray}
\begin{eqnarray}
\; \chi_{\text{loc}}^{132}&  = & \sum_{N,M,L,S}\frac{1}{i(\np+\o)-\EM+\EL}\left[
\frac{1}{i(\nu+\o)+\EL-\ES}
 \left(\frac{\ExN+\ExS}{i\np-\EN+\ES}+\frac{\ExL-\ExN}{i(\n+\np+\o)-\EN+\EL}\right) \nonumber \right. \\ & + & \left. \frac{1}{i(\n-\np)+\EM-\ES}
\left(\frac{\ExM+\ExN}{i\n+\EM-\EN}-\frac{\ExS+\ExN}{i\np+\ES-\EN}\right)\right]\times\cd_{NM}\cpd_{ML}\c_{LS}\cp_{SN}
\end{eqnarray}
\begin{eqnarray}
\chi_{\text{loc}}^{213}&  = & \sum_{N,M,L,S}\frac{1}{i(\np+\o)+\EL-\EM}
\frac{1}{i(\n+\o)+\ES-\EN}\left(\frac{\ExM-\ExS}{i\o+\ES-\EM}+
\frac{\ExL-\ExN}{i(\n+\np+\o)-\EN+\EL}\right. \nonumber \\ & + & \left.
\frac{\ExM+\ExN}{i\n+\EM-\EN}-\frac{\ExS+\ExL}{i\np+\EL-\ES}\right)
\times\cd_{NM}\cpd_{ML}\cp_{LS}\c_{SN}
\end{eqnarray}
\begin{eqnarray}
\; \chi_{\text{loc}}^{231}&  = & \sum_{N,M,L,S}\frac{-1}{i(\np+\o)+\EM-\EN}\left[
\frac{1}{i(\n-\np)-\EM+\ES} \left(\frac{\ExM+\ExL}{i\n+\EL-\EM}-
\frac{\ExL+\ExS}{i\np+\EL-\ES}\right) \nonumber \right. \\ & - & \left.
\frac{1}{i(\n+\o)+\ES-\EN}  \left(\frac{\ExL-\ExN}{i(\n+\np+\o)+\EL-\EN}-
\frac{\ExL+\ExS}{i\np+\EL-\ES}\right)\right]\times\cpd_{NM}\cd_{ML}\cp_{LS}\c_{SN} \nonumber \\
\end{eqnarray}
 \begin{eqnarray}
\chi_{\text{loc}}^{312} &  = & \sum_{N,M,L,S}\frac{-1}{i(\np+\o)+\EM-\EN}
\frac{1}{i(\n+\o)+\EL-\ES}\left(\frac{\ExM-\ExS}{i\o+\EM-\ES}+
\frac{\ExL+\ExM}{i\n+\EL-\EM}\right. \nonumber \\ & - & \left.
\frac{\ExN+\ExS}{i\np+\ES-\EN}+\frac{\ExN-\ExL}{i(\n+\np+\o)+\EL-\EN}\right)
\times\cpd_{NM}\cd_{ML}\c_{LS}\cp_{SN}
\end{eqnarray}
\begin{eqnarray}
\label{Eq:chi321}
\chi_{\text{loc}}^{321} &  = & \sum_{N,M,L,S}\frac{1}{i(\np+\o)+\EM-\EN}\left[
\frac{1}{i(\n+\o)+\EM-\EL}
 \left(\frac{\ExS+\ExL}{i\n+\ES-\EL}+\frac{\ExS-\ExM}{i\o+\EM-\ES}\right) \nonumber \right. \\ & - & \left. \frac{1}{i(\n-\np)+\EN-\EL}
\left(\frac{\ExS+\ExL}{i\nu+\ES-\EL}-\frac{\ExS+\ExN}{i\np+\ES-\EN}\right)\right]\times\cpd_{NM}\c_{ML}\cd_{LS}\cp_{SN}
\end{eqnarray}
\end{widetext}

After the DMFT self-consistency condition has been fulfilled, the ED-DMFT
evaluation of the local susceptibility Eq. (\ref{Eq:chiloc})
is  obtained directly by plugging the eigenvalues $E_N$ and
the matrix elements $\c_{NM}$  of the associated AIM in the Eqs.
(\ref{Eq:chi123})-(\ref{Eq:chi321}) and performing the corresponding four
summations over the Hilbert space. The relevant numerical effort related to
the Hilbert space summations can be handled  by means of a parallel
computations for the case considered here (i.e., $N_s=5$ and 
$N_{max} \ge 20$).
States for which all Boltzmann weights ($e^{-\beta E}$) or
matrix elements are smaller than $10^{-6}$ are neglected.

For the numerical implementation, let us note that some denominators
in Eqs. (\ref{Eq:chi123})-(\ref{Eq:chi321}), characterized
by a bosonic Matsubara frequency  (e.g., those with $i\o$ or
with $i(\n-\np)$), can vanish during the summations over the Hilbert space.

However there are no divergences, since the corresponding
numerators are simultaneously vanishing, so that their limit is always well
defined.
To avoid computational problems, we simply add a very small energy
shift  in all the terms with a
vanishing denominator (e.g., $E_N \rightarrow E_N +10^{-8}$),
in order to evaluate numerically the correct limiting values.


\begin{references}
\bibitem{note_RMP} This three frequency vertex has been already
introduced in Ref.\ \onlinecite{DMFT2},
  but, to our knowledge, only the static ($\omega =0$) vertex has been
calculated, and used for computing DMFT static susceptibilities.
\bibitem{note_proof} Let us note that the analytical continuation sometimes 
also shows a little bump for the U=2D case. But quantitatively the results
are very similar.
\bibitem{note_af} The generalization of the D$\Gamma$A to the case of 
magnetically ordered phases is straightforward, albeit lengthy:
 it is necessary to consider different vertices for
the longitudinal and transverse spin fluctuations.
\end{references}
\end{document}